\documentclass[a4paper,11pt]{article}
\usepackage{jheppub}
\usepackage{comment}
\usepackage{amsmath,amssymb,extarrows,mathtools,graphicx,subfigure,setspace,bbold}
\usepackage{tikz}
\usetikzlibrary{decorations.pathreplacing}
\newcommand{\be}{\begin{equation}}
\newcommand{\bea}{\begin{eqnarray}}
\newcommand{\eea}{\end{eqnarray}}
\newcommand{\ba}{\begin{array}}
\newcommand{\ea}{\end{array}}
\newcommand{\bal}{\begin{align}}
\newcommand{\eal}{\end{align}}
\newcommand{\ee}{\end{equation}}
\newcommand{\bes}{\begin{equation*}}
\newcommand{\beas}{\begin{eqnarray*}}
\newcommand{\eeas}{\end{eqnarray*}}
\newcommand{\bas}{\begin{array*}}
\newcommand{\eas}{\end{array*}}
\newcommand{\ees}{\end{equation*}}

\newcommand{\p}{\partial}
\title{Crystalline Geometries from Fermionic Vortex Lattice}
\author{M. Reza Mohammadi Mozaffar,}
\author{Ali Mollabashi}

\affiliation{School of physics, Institute for Research in Fundamental Sciences (IPM), Tehran, Iran}
\emailAdd{m$_{-}$mohammadi@ipm.ir}
\emailAdd{mollabashi@ipm.ir}

\abstract{We study charged Dirac fermions on an AdS$_2\times \mathbb{R}^2$ background with a non-zero magnetic field. Under certain boundary conditions, we show that the charged fermion can make the background unstable, resulting in spontaneously formation of a vortex lattice.
We observe that an electric field emerges in the back-reacted solution due to the vortex lattice constructed from spin polarized fermions. This electric field may be extended to the UV boundary which leads to a finite charge density. We also discuss corrections to the thermodynamic functions due to the lattice formation.}

\begin{document}
\begin{flushright} \small
IPM/P-2013/020
\end{flushright}
\maketitle
\flushbottom
\section{Introduction}
AdS/CFT correspondence has provided a powerful tool for studying strongly coupled field theories. This correspondence has been applied to certain systems in condensed matter physics (for excellent reviews see \cite{Iqbal:2011ae, Hartnoll:2009sz, Hartnoll:2011fn}). Specifically in the last few years, much attempts have been devoted to apply this approach to various phenomena including superconductivity, Fermi and non-Fermi liquids.
From gravity point of view, this can be done by considering different matter fields on an AdS-RN geometry which corresponds to physical systems at finite temperature and finite density.  
A natural question one may pose is how to consider the effects of the underlying lattice that strongly coupled systems leave on. This key ingredient of condensed matter systems was neglected in most of the earlier studies (see however \cite{Kachru:2009xf,Kachru:2010dk,Hellerman:2002qa}).

To consider a lattice, one should break translational invariance of the dual field theory. This was first motivated by the significant effect of the momentum dissipation of charge carriers in optical conductivity \cite{Horowitz:2012ky,Horowitz:2012gs}. In these papers the authors have explicitly broken the translational symmetry in two different ways. Either by imposing a spatially inhomogeneous periodic source for a neutral scalar field coupled to an Einstein-Maxwell theory \cite{Horowitz:2012ky}, or alternatively by considering the back-reaction of a periodic chemical potential on the metric in an Einstein-Maxwell theory \cite{Horowitz:2012gs}. In both cases, lattice effects were handled by solving coupled PDE's numerically. The most important achievement of these models was holographic reconstruction of Drude peak in optical conductivity and the power-law behavior in an intermediate frequency range\footnote{Fermions on these backgrounds have been studied in \cite{Ling:2013aya}.}.

Other approaches have also been used to study lattice effects in the dual field theory. An explicit example 
is five dimensional models deformed by a uniform chemical potential. Although translational symmetry is broken in these models, a Bianchi VII$_0$ subgroup is still preserved giving the dual field theory a helical structure \cite{Nakamura:2009tf,Donos:2011ff}. These models are technically much easier to deal with since the homogeneity of the system at constant $r$ slices leads to ODE's rather than PDE's \cite{Donos:2011ff,Donos:2012wi}\footnote{The story is different in four dimensions, see for example \cite{Donos:2013wia}.}. Ground states of these models that the spatial modulation persists deep in the IR, are constructed numerically in \cite{Donos:2011ff,Donos:2012js}.

Beside the above approaches, recently an analytical back-reacted crystalline geometry was constructed in \cite{Bao:2013fda}. The gravity dual of this model is an AdS$_2\times \mathbb{R}^2$ supported by a magnetic field, which breaks translational symmetry. The vortex lattice is constructed via the instability of a probe charged scalar field coupled to the magnetic field.	The important distinction between incorporating the lattice by means of periodic chemical potential \cite{Horowitz:2012ky,Horowitz:2012gs} and this solution is their behavior at IR. While in the former case, the background charge carriers screen the spatially modulated chemical potential in the IR, in the latter case, since magnetic field can not be screened, the effect of the lattice could persist deep in the IR. This solution has also been generalized to gravity duals with Lifshitz and/or hyperscaling violation exponents \cite{Bao:2013ixa}. It is worth noting that this class of solutions is based on an elegant vortex lattice solution constructed in \cite{Maeda:2009vf,Albash:2008eh}\footnote{See also \cite{Bu:2012mq} where the back-reaction of such a lattice on the gauge filed is studied.}.

In this paper we consider the same background as that in \cite{Bao:2013fda}, though in the present case the background is probed by a charged Dirac fermion. We construct a fermionic vortex lattice by means of the lowest Landau level (LLL) solutions. Fermionic LLL states are spin polarized and their holographic aspects have been previously discussed in \cite{Blake:2012tp,Bolognesi:2012pi}. In the present work we are interested in the spontaneous formation of a crystalline geometry sourced by these LLLs on the gauge sector. Thus we will have to analytically solve the corresponding coupled PDE's for the metric and the gauge field to the lowest non-trivial order. We show that the back-reaction of the fermionic lattice leads to an emergent electric field and thus an effective charge density. In a specific range of parameters we show that the electric field can reach the UV boundary. The situation is different from \cite{Bao:2013fda}, where a lattice structure due to a charged scalar condensate, only corrects the back-ground magnetic field.

The rest of this paper is organized in the following way. Section 2 is devoted to the basic setup including the gravitational background, Dirac action and its possible boundary terms. In Section 3 we will consider a charge fermion on an  AdS$_2\times \mathbb{R}^2$ background where we will construct the fermionic lattice from LLL solution. In the rest of this section we will study the back-reaction on the metric and gauge field sourced by these fermions. In the last section we discuss about our results and directions for further investigation of the model. 

\section{Basic set-up}
In this section we describe basic ingredients for constructing a fermionic lattice. In Section (\ref{gbg}) we will review the gravitational back-ground. Since dealing with fermions in curved space-time requires special care, we shall review certain basic properties of fermions in curved space-times in subsection (\ref{frm}). In particular, in order to provide a full description of dynamics, one must add suitable boundary terms to the Dirac action \cite{Henneaux:1998ch, Iqbal:2009fd}. We will discuss different boundary conditions which lead to well-defined variational principle and also lead to lattice formation.
\subsection{Gravitational back-ground}\label{gbg}
Consider the following action which may support a magnetic AdS$_2\times \mathbb{R}^2$ solution
\begin{eqnarray}\label{action1}
S_{1}=\int{d^4x\sqrt{-g}\left(R-\frac{1}{4}F_{\mu \nu}F^{\mu \nu}-2\Lambda\right)}.
\end{eqnarray}
The corresponding metric and gauge field are
\begin{eqnarray}\label{background}
ds^2=L^2\left(-\frac{dt^2}{r^2}+\frac{dr^2}{r^2}+dx^2+dy^2\right),\hspace*{1cm} F=B dx\wedge dy,
\end{eqnarray}
where $L^{-2}=-2\Lambda>0$ and $r\to0$ is the UV boundary. The equations of motion fixes $B$ in terms of the AdS radius as
\beas
B=\sqrt{2}L.
\eeas
So this solution has one free parameter, which can be considered as the magnetic field.

It is worth noting that this solution emerges as the near-horizon limit of magnetically charged extremal AdS-RN black-branes. The holographic dual is an emergent IR CFT which describes the semi-local quantum liquid phase that plays a key role in explaining non-Fermi liquid behaviours and quantum phase transitions (see \cite{Iqbal:2011ae} for details).
\subsection{Dirac action and boundary terms}\label{frm}
In order to study the vortex lattice solution with a charged fermionic probe, we add the following Dirac action to the back-ground \eqref{action1}
\begin{eqnarray}\label{action2}
S_{2}=\int{d^4x\sqrt{-g}\,i\,\overline{\Psi} \left( \frac{1}{2}(\Gamma^{\mu} \overrightarrow{D}_{\mu}- \overleftarrow{D}_{\mu}\Gamma^{\mu})-m\right) \Psi}.
\end{eqnarray}
Here $\Gamma^{\mu}D_{\mu} \equiv e_a^{\mu}\Gamma^{a}\left(\partial_{\mu}+\frac{1}{8}\omega_{ab,\mu}[\Gamma^a,\Gamma^b]+iqA_{\mu}\right)$ is the covariant derivative where $\omega_{ab,\mu}$ is the spin connection and the vierbein $e_a^{\mu}$ translates between space-time indices $\mu, \nu$ and tangent space indices $a, b$.  The gamma matrices carry tangent space indices and obey the Clifford algebra $\{\Gamma^a,\Gamma^b\}=2\eta^{ab}$. We define the chiral gamma matrix as $\Gamma^5=i\Gamma^r\Gamma^t\Gamma^x\Gamma^y$ and conjugate spinors are defined by $\overline{\Psi}=\Psi^{\dagger} \Gamma^t$. $q$ denotes the electric charge of the fermion which we will set it to unity without loss of generality. 

The magnetic field in the $x$-direction forces us to decompose the Dirac spinor eigenvectors of $i\Gamma^x\Gamma^y$ operator. These eigenvectors also correspond to spin up and spin down states. Thus we use the projection operator
\begin{eqnarray}\label{leftright}
\Psi_{\pm}=P_{\pm}\Psi, \hspace*{1cm} P_{\pm}=\frac{1}{2}\left(1\pm i \Gamma^x \Gamma^y\right).
\end{eqnarray}
A suitable basis which we choose for the Dirac matrices is
\begin{eqnarray}\label{basis}
\Gamma^x=\left(\begin{array}{cc}
\sigma^2 &0\\
0&-\sigma^2
\end{array}\right),\Gamma^y=\left(
\begin{array}{cc}
 \sigma ^1 & 0 \\
 0 & \sigma ^1
\end{array}
\right),\Gamma ^t=\left(
\begin{array}{cc}
 -i \sigma ^3 & 0 \\
 0 & -i \sigma ^3
\end{array}
\right),\Gamma ^r=\left(
\begin{array}{cc}
 0 & -i \sigma ^2 \\
 i \sigma ^2 & 0
\end{array}
\right),
\end{eqnarray}
which implies the form of spin down and spin up spinors as
\begin{eqnarray}\label{spinor}
\Psi^-=\left(
\begin{array}{c}
 0 \\
 {\Psi_1^-}\\
 {\Psi_2^-}\\
 0
\end{array}
\right)
\;\;\;,\;\;\;\Psi^+=\left(
\begin{array}{c}
 {\Psi_1^+} \\
 0\\
 0\\
 {\Psi_2^+}
\end{array}
\right).
\end{eqnarray}

Since we are interested in finding a normalizable solution for the fermion in the entire bulk geometry, in order to obtain a non-trivial solution we have to terminate the geometry in the IR. This could be done by considering a black-brane horizon or a hard wall. In what follows, we will impose a hard wall; a wall that abruptly cuts the geometry at some finite $r=r_0$. As mentioned before, imposing boundary conditions for fermions is not a trivial task. The rest of this section is devoted to these subtleties.

Standard and alternative boundary conditions for fermions can be imposed by 
\begin{eqnarray}\label{standard}
(1\mp \Gamma^r)\Psi(r\to0)=0.
\end{eqnarray}
In order to have a well-defined variational principle with this sort of boundary conditions, we must add a boundary term to action \eqref{action2} as follows
\begin{eqnarray}\label{standard}
S_{\mathrm{bdy}}^{\mathrm{UV}}=\pm \frac{i}{2}\int_{r=0}{d^3x\sqrt{-h}\,\overline{\Psi}\Psi},
\end{eqnarray}
where the upper and lower signs refer to standard and alternative quantizations and $h$ is the determinant of the induced boundary metric, $h=gg^{rr}$. It is important to notice that the alternative quantization is allowed in a specific range of the spinor mass. For the LLL solutions which will be discussed in Sec. (\ref{3.2}) this range is $0<mL<1/2$.

Since we are interested in spontaneous formation of the lattice, we will turn off the source of the fermionic field. In order to read the source in our basis, we have to consider the variation of the full Dirac action, which leads to
\bea\label{varbulk}
\delta S_{2}=\text{bulk term}&+&\frac{i}{2}\int_{r=0}{d^3x\sqrt{-h}\left(\delta\overline{\Psi}\Gamma^r\Psi-\overline{\Psi}\Gamma^r \delta\Psi\right)}\nonumber\\&-&\frac{i}{2}\int_{r=r_0}{d^3x\sqrt{-h}\left(\delta\overline{\Psi}\Gamma^r\Psi-\overline{\Psi}\Gamma^r \delta\Psi\right)}.
\eea
First we consider the UV boundary term for the standard quantization, which after variation becomes
\begin{eqnarray}\label{uvbdy}
\delta S_{\mathrm{bdy}}^{\mathrm{UV}}=\frac{i}{2}\int_{r=0}{d^3x\sqrt{-h}\left(\delta\overline{\Psi}\Psi+\overline{\Psi}\delta\Psi\right)}.
\end{eqnarray}
Thus the variation of the full action at the UV boundary reads
\begin{eqnarray}\label{total}
{\delta S_{2}}\Big|_{r=0}+\delta S_{\mathrm{bdy}}^{\mathrm{UV}}=\frac{1}{2}\int_{r=0}{d^3x\sqrt{-h}\left(\delta\xi_+^\dagger \xi_-+\xi_-^\dagger \delta\xi_+-\delta\chi_-^\dagger \chi_+-\chi_+^\dagger \delta\chi_-\right)},
\end{eqnarray}
where $\xi_\pm=\Psi_1^+\pm \Psi_2^+$ and $\chi_\pm=\Psi_1^-\pm \Psi_2^-$. Thus the UV boundary condition for our choice becomes
\begin{eqnarray}\label{sourcestandard}
\xi_+=0,\hspace*{1cm}\chi_-=0.
\end{eqnarray}
For the alternative quantization these combinations become $\xi_-=0$ and $\chi_+=0$.

For a boundary condition at IR boundary, we will follow a procedure similar to \cite{Bolognesi:2012pi} and add the following boundary term to the action \eqref{action2}
\begin{eqnarray}\label{irbdy}
S_{\mathrm{bdy}}^{\mathrm{IR}}=-\frac{i}{2}\int_{r=r_0}{d^3x\sqrt{-h}\,\overline{\Psi}e^{i\left(\theta-\frac{\pi}{2}\right)\Gamma^5}\Psi}.
\end{eqnarray}
So that
\begin{eqnarray}\label{varirbdy}
{\delta S_{2}}\Big|_{r=r_0}+\delta S_{\mathrm{bdy}}^{\mathrm{IR}}=-\int_{r=r_0}{d^3x\sqrt{-h} \left(\delta\tilde{\xi}_-^\dagger \tilde{\xi}_++\tilde{\xi}_+^\dagger \delta\tilde{\xi}_-+\delta\tilde{\chi}_+^\dagger\tilde{\chi}_-+\tilde{\chi}_-^\dagger \delta\tilde{\chi}_+\right)},
\end{eqnarray}
where
\begin{eqnarray}
\left(\begin{array}{c}
\tilde{\xi}_+\\
\tilde{\xi}_-
\end{array}\right)
=\left(\begin{array}{cc}
 \cos\frac{\theta}{2}& \sin\frac{\theta}{2}\\
 \sin\frac{\theta}{2}& -\cos\frac{\theta}{2}
\end{array}\right) \left(\begin{array}{c}
\Psi^+_1\\
 \Psi^+_2
\end{array}\right),\hspace*{0.5cm}
\left(\begin{array}{c}
\tilde{\chi}_+\\
\tilde{\chi}_-
\end{array}\right)
=\left(\begin{array}{cc}
 -\cos\frac{\theta}{2}& \sin\frac{\theta}{2}\\
 \sin\frac{\theta}{2}& \cos\frac{\theta}{2}
\end{array}\right) \left(\begin{array}{c}
\Psi^-_1\\
 \Psi^-_2
\end{array}\right).
\end{eqnarray}
Therefore a well-defined variational principle is obtained on the hard wall by choosing
\begin{eqnarray}\label{irsource}
\tilde{\xi}_-=0,\hspace*{1cm}\tilde{\chi}_+=0.
\end{eqnarray}
We will show that these two boundary conditions at UV and IR \eqref{sourcestandard}-\eqref{irsource} lead to a unique normalizable solution for the Dirac hair. For the LLL solution discussed in subsection (\ref{3.2}), the condition \eqref{irsource} is satisfied by choosing $\theta=\pi/2$ for standard and $\theta=-\pi/2$ for alternative quantizations.

Note that the IR boundary condition we have used on the hard wall is completely different from what is used in \cite{Bao:2013fda}. This is because imposing Dirichlet or Neumann boundary conditions on the normalizable mode of a scalar filed at the IR boundary just yields to the trivial solution of AdS$_2 \times \mathbb{R}^2$. Thus authors of \cite{Bao:2013fda} have imposed a Randall-Sundrum like (see \cite{Randall:1999ee}) boundary condition on a hard wall which supports a non-trivial profile for the scalar field\footnote{We thank Ning Bao for bringing our attention to this point.}. The prescription is to add a mirror image of the space-time to the other side of the wall and glue them together at the IR boundary. So the set-up has two UV boundaries and each field also has a mirror image, which reduces to the desired space-time after imposing a Z$_2$ symmetry.
This implies a discontinuity in the first derivative of the fields at the wall, although the fields are continuous there.

In our set-up, as we have seen, it is not necessary to consider a mirror boundary condition to obtain a non-trivial fermionic profile, though it is still possible to do that. Again imposing such a boundary condition for fermions is accompanied by some complications (for some early ideas see \cite{Flachi:2001bj}). One must consider spinors  that are representations of Z$_2$ group. Imposing the Z$_2$ invariance on the Dirac action \eqref{action2} leads to 
\begin{eqnarray}\label{z2}
\Psi(-r,x_a)=\pm \Gamma^r \Psi(r,x_a),
\end{eqnarray}
which is the suitable boundary condition in a mirror geometry for a fermionic field.

\section{The crystalline geometry}
In this section we find the IR instability due to a fermionic probe which leads to a crystalline ground state. We will see that by changing the parameters, the $\Psi=0$ solution can be degenerated with a vortex lattice solution. The onset of the instability, referred by the critical point, is identified by the existence of a normalizable solution for $\Psi$ that satisfies both the UV and IR boundary conditions. After finding the desired solution for a Dirac hair \cite{Cubrovic:2010bf}, we will solve the full set of equations of motion, including the back-reaction of the lattice on the gauge sector.

The equations of motion are
\begin{eqnarray}\label{eoms}
(\Gamma^{\mu}D_{\mu}-m)\Psi &=&0\\
\frac{1}{\sqrt{-g}}\partial_\mu\left({\sqrt{-g}F^{\mu \nu}}\right)&=&J^\nu_{\Psi}\nonumber\\
G_{\mu \nu}+\Lambda g_{\mu \nu}&=&T^{A}_{\mu \nu}+T^{\Psi}_{\mu \nu}\nonumber,
\end{eqnarray}
where
\begin{eqnarray}\label{current}
J^\nu_{\Psi}&=&\overline{\Psi}\Gamma^\nu \Psi\\
T^{A}_{\mu \nu}&=&\frac{1}{2}F_{\mu \lambda}F_{\nu}^{\lambda}-\frac{1}{8}g_{\mu \nu}F^2\nonumber\\
T^{\Psi}_{\mu \nu}&=&-\frac{i}{8}\left\{\overline{\Psi} e_{a\mu}\left(\partial_\nu+\frac{1}{4}\omega _{bc,\nu}\Gamma^{bc}+iA_\nu\right)\Gamma^a \Psi+\text{h.c.}\right\}+(\mu \leftrightarrow\nu)\nonumber.
\end{eqnarray}

In order to consider the back-reaction of the fermionic lattice on the back-ground \eqref{background}, we consider a perturbative expansion around the critical point. At such a point the value of the fermionic field is zero, so one can consider
\begin{eqnarray}\label{fermionexpand}
\Psi(r,x,y)=\epsilon \Psi_{(1)}(r,x,y)+\epsilon^3 \Psi_{(3)}(r,x,y)+\cdots.
\end{eqnarray}
The expansion parameter $\epsilon$ is the distance away from the critical point in the parameter space. We are interested in the back-reaction of the LLL solutions (Sec. \ref{3.2}) to the gauge sector. A simple analysis shows that the only non-trivial sources at order  $\mathcal{O}(\epsilon^2)$ are $T^{\Psi}_{tx}$, $T^{\Psi}_{ty}$, and $J_{\Psi}^{t}$. Thus we will use the following ansatz for the back-reacted metric and gauge field
\begin{eqnarray}\label{gaugeexpand}
ds^2&=&L^2\left[-\frac{dt^2}{r^2}+\frac{dr^2}{r^2}+\epsilon^2 \Big(a(r,x,y)dtdx+b(r,x,y)dtdy\Big)+dx^2+dy^2\right],\nonumber\\
A&=&By dx+\epsilon^2 a_2^t(r,x,y) dt.
\end{eqnarray}
This shows that the back-reaction of the fermionic lattice on the gauge sector at the leading order leads to an effective charge density. The situation is different from \cite{Bao:2013fda}, where a lattice structure due to a scalar condensate just corrects the back-ground magnetic field.

\subsection{Droplet solution}
At order $\epsilon$, we can neglect the back-reaction of $\Psi_{(1)}$ on the gauge sector\footnote{From here we omit the subscript (1) in $\Psi_{(1)}$.}. In this limit from the equations in \eqref{eoms}, only the Dirac equation is relevant. However, when dealing with fermions in a curved space-time, it is often more simple to get rid of spin connection terms by introducing a rescaled fermionic field $\Psi(r,x,y)=(-h)^{-1/4}\,\psi(r,x,y)$. Making use of this, the Dirac equation on the back-ground \eqref{background} takes the following form
\begin{eqnarray}\label{first}
\left(\Gamma^r r \partial_r+\Gamma^x\left(\partial_x +iBy\right)+\Gamma^y \partial_y-mL \right) \psi=0.
\end{eqnarray}
By acting $(\Gamma^{\mu}D_{\mu}+mL)$ operator on the above equation and after some gymnastics with gamma matrices, one can find a second order equation as follows
\begin{eqnarray}\label{second}
\left[r^2\p_r^2+\p_x^2+\p_y^2+r\p_r+2iBy\p_x-B^2y^2-iB\Gamma^x\Gamma^y-m^2L^2\right]\psi(r,x,y)=0.
\end{eqnarray}
We will solve the above equation by separation of variables as $\psi_{\pm}(r,x,y)=\rho(r)g(y)e^{ikx}C_{\pm}$ where $C_{\pm}$ are constant spinors such that $i\Gamma^x\Gamma^y C_{\pm}=\pm C_{\pm}$. The separated equations become
\begin{eqnarray}\label{separation}
r^2\frac{\rho_{n_{\pm}}''}{\rho_{n_{\pm}}}+r\frac{\rho_{n_{\pm}}'}{\rho_{n_{\pm}}}-m^2L^2=(k+By)^2-\frac{g_{n_{\pm}}''}{g_{n_{\pm}}}\pm B=-\lambda_{n_{\pm}},
\end{eqnarray}
where $\lambda_{n_{\pm}}$ are the eigenvalues from the separation of variables. To find the 'basic droplet' solution, it is enough to consider the equation for $g(y)$, setting $Y=\sqrt{B}(y+\frac{k}{B})$, one finds
\beas
g''_{n_{\pm}}(Y)-g_{n_{\pm}}(Y)\left(Y^2+\frac{\lambda_{n_{\pm}}}{B}\pm 1\right)=0.
\eeas
A general solution of the above equation is the parabolic cylinder function, though if one demands a normalizable solution as $y\rightarrow\infty$, then it becomes the familiar Hermite function\footnote{Assuming $B>0$.}
\beas
g_{n_{\pm}}(Y)\sim e^{\frac{-Y^2}{2}}H_{n_{\pm}}(Y)
\eeas
where $c^{\pm}$'s are constants, and $\lambda_{n_{\pm}}=-2B(n_{\pm}+\frac{1}{2}\pm\frac{1}{2})$. Actually the situation is the same as a quantum harmonic oscillator eigenvalue problem and the corresponding Landau levels, but here for a fermionic field. Various aspects of these solutions have been previously discussed in the series of papers \cite{Albash:2009wz,Albash:2010yr,Gubankova:2010rc,Bolognesi:2011un,Bolognesi:2012pi,Blake:2012tp}, in a related but distinct context. 

Now we will solve the radial part of the equation \eqref{separation} for $\rho(r)$. It has a power law solution in AdS$_2 \times \mathbb{R}^2$ as
\beas
\rho_{n_{\pm}}(r)=c^{\pm} r^{\alpha_{\pm}}+d^{\pm} r^{-\alpha_{\pm}}, \hspace*{1cm}\alpha_{\pm}=\sqrt{m^2L^2-\lambda_{n_{\pm}}}.
\eeas
So the full solution of the equation \eqref{second} becomes
\beas
\psi(r,x,Y)=e^{ikx}e^{\frac{-Y^2}{2}}\left[\left(c^{+} r^{\alpha_{+}}+d^{+} r^{-\alpha_{+}}\right)H_{n_{+}}(Y)+\left(c^{-} r^{\alpha_{-}}+d^{-} r^{-\alpha_{-}}\right)H_{n_{-}}(Y)\right]
\eeas
where $c^{\pm}$ and $d^{\pm}$ are constant spinors. Since the Dirac equation \eqref{first} is a first order equation, $c^{\pm}$ and $d^{\pm}$ are not independent constants. In other words the desired solutions are those which satisfy the original first order equation \eqref{first} among the above solutions. This can be done, by using the recursion relations between the Hermite functions, it becomes obvious that the existence of non-trivial solution implies that the $\alpha_+=\alpha_-\equiv\alpha$ and thus $\lambda_{n_+}=\lambda_{n_-}\equiv\lambda$. This leads to
\beas
n_-=n_++1\equiv n.
\eeas
Using these constrains, the relations between the constant spinors (for $n\ne0$) becomes
\begin{eqnarray}\label{cd}
{c_1^+}=\frac{1}{2n}\left(\nu c_1^-+\sqrt{\nu^2+2n} c_2^-\right),\hspace*{0.5cm}{c_2^+}=\frac{1}{2n}\left(\nu c_2^-+\sqrt{\nu^2+2n} c_1^-\right)\nonumber\\
{d_1^+}=\frac{1}{2n}\left(\nu d_1^--\sqrt{\nu^2+2n} d_2^-\right),\hspace*{0.5cm}{d_2^+}=\frac{1}{2n}\left(\nu d_2^--\sqrt{\nu^2+2n} d_1^-\right)
\end{eqnarray}
where $\nu^2=\frac{m^2L^2}{B}$ and $\alpha=\sqrt{m^2L^2+2nB}$. Thus the physical solution can be written in terms of Hermite polynomials as follows
\begin{eqnarray}\label{PSI}
\Psi(r,x,Y) =r^{\alpha+\frac{1}{2}}e^{ikx}e^{\frac{-Y^2}{2}}\left[\left(c^-+d^-r^{-2\alpha}\right) H_{n}(Y)+\left(c^++d^+r^{-2\alpha}\right) H_{n-1}(Y)\right].
\end{eqnarray}

As we have mentioned earlier, in order to have a crystalline ground-state, we must have normalizable solution for $\Psi$ that satisfies the IR boundary conditions. The above solution have two different radial modes which one of them can diverge near the boundary at $r=0$ depending on the parameters. In the case that $\alpha>\frac{1}{2}$, we can only consider the standard quantization, but in the case that $\alpha<\frac{1}{2}$, the alternative quantization is also possible.
\subsection{Fermionic vortex lattice}\label{3.2}
As mentioned in \cite{Maeda:2009vf}, to obtain the vortex lattice structure from the single droplet solution, it is enough to consider the $n=0$ level. It is evident from the equation \eqref{cd} that in this case the general solution \eqref{PSI} become useless. Actually in this case the $H_{-1}$, which is the eigenfunction of the spin up fermion is not well-defined  and one must define $c_+=d_+\equiv0$. After some simple algebra one finds
\begin{eqnarray}\label{PSI0}
\Psi_0(r,x,y) =r^{\frac{1}{2}}e^{ikx}\psi_0(y;k)\left(
\begin{array}{c}
0 \\
 c_0 r^{mL}+d_0 r^{-mL}\\
 c_0 r^{mL}-d_0 r^{-mL}\\
0
\end{array}
\right),\hspace*{1cm}\psi_0(y;k)=e^{-\frac{B}{2}(y+\frac{k}{B})^2},
\end{eqnarray}
where the Hermite function is normalized such that $H_0=1$. So the lowest Landau level is spin polarized and the degeneracy of fermions is half of the higher levels. The vortex lattice solution can be obtain by an appropriate superposition of the droplet solutions
\begin{eqnarray}\label{lat}
\Psi_0^{\mathrm{lat}}(x,y)=\sum\limits_{l=-\infty}^\infty c_l e^{ik_l x}\psi_0(y;k_l)
\end{eqnarray}
where 
\begin{eqnarray}
c_l\equiv e^{-i\pi\frac{v_2}{v_1^2}l^2},\hspace*{1cm}k_l=\frac{2\pi l}{v_1}\sqrt{B}
\end{eqnarray}
for arbitrary $v_1$ and $v_2$. In terms of the elliptic theta function $\vartheta_3$ defined by
\begin{eqnarray}
\vartheta_3(v,\tau)\equiv \sum\limits_{l=-\infty}^\infty q^{l^2}z^{2l}, \hspace*{1cm}q \equiv e^{i\pi \tau},\hspace*{1cm}z \equiv e^{i\pi v}
\end{eqnarray}
the equation \eqref{lat} becomes
\begin{eqnarray}\label{lat1}
\Psi_0^{\mathrm{lat}}(x,y)=e^{-\frac{By^2}{2}}\vartheta_3(v,\tau)
\end{eqnarray}
where
\begin{eqnarray}
v=\frac{\sqrt{B}(x+iy)}{v_1}, \hspace*{1cm}\tau=\frac{2\pi i-v_2}{v_1^2}.
\end{eqnarray}
The elliptic theta function $\vartheta_3$ has two properties which implies the vortex lattice structure.
The first one is its pseudo-periodicity
\begin{eqnarray}\label{pseudo}
\vartheta_3(v+1,\tau)=\vartheta_3(v,\tau), \hspace*{1cm}\vartheta_3(v+\tau,\tau)=e^{-2\pi i(v+\tau/2)}\vartheta_3(v,\tau),
\end{eqnarray}
thus every function that depends on the norm of $\vartheta_3$ is invariant upon translation by the lattice generators
\begin{eqnarray}\label{latticegen}
\mathbf{b}_1=\frac{1}{\sqrt{B}}v_1\partial_x,\hspace*{1,cm}\mathbf{b}_2=\frac{1}{\sqrt{B}}\left(\frac{2\pi}{v_1}\partial_y+\frac{v_2}{v_1}\partial_x\right).
\end{eqnarray}
By this choice, every unit cell contains exactly one quantum flux, where the area is given by $2\pi/B$. Second, $\vartheta_3$ vanishes at
\begin{eqnarray}\label{core}
\mathbf{x}_{m,n}=\left(m+\frac{1}{2}\right)\mathbf{b}_1+\left(n+\frac{1}{2}\right)\mathbf{b}_2, \hspace*{1cm}m,n \in \mathbb{N},
\end{eqnarray}
and has a phase rotation of $2\pi$ around each such zero, thus one can consider $\mathbf{x}_{m,n}$ as the vortex cores. By changing the parameters $v_1$ and $v_2$, one can construct various lattice shapes, such as rectangular, square, rhombic, and etc. In this paper we will only consider the square lattice which is obtained by setting 
\begin{eqnarray}\label{rectangular}
v_2=0\rightarrow c_l=1\hspace*{1cm}\text{and}\hspace*{1cm}v_1=\sqrt{2\pi}.
\end{eqnarray}
Now that we are equipped with the fermionic vortex lattice, constructed from lowest Landau level of a Dirac fermion, we can consider the back-reaction of this lattice structure on the gauge sector.
\subsection{Back-reaction on the gauge sector}
The back-reaction of the crystalline structure on the metric and the gauge field at order $\mathcal{O}(\epsilon^2)$ is sourced by the fermions as matter current and energy-momentum tensor. As we mentioned earlier, in order to obtain a spontaneous lattice formation, the source must be turned off. This means that for the standard quantization, one must consider $d_0=0$ in the solution \eqref{PSI0}, where we used the equation \eqref{sourcestandard} for identifying the source term\footnote{The following calculation can be held in a similar way for the case of alternative quantization. This can be done by changing: $c_0 \rightarrow d_0, m\rightarrow -m.$}. Dealing with the equations is much simpler if we extract the $r$ scaling in the $\mathcal{O}(\epsilon^2)$ corrections and solve the equations for the spatial dependence. We assume
\begin{eqnarray}\label{correct}
f_i(r,x,y)=r^{2m L}f_i(x,y)
\end{eqnarray}
where $f_i=a,b,a_2^t$.
At this order, the only non-trivial Maxwell equation is
\begin{eqnarray}\label{backeommax}
(\partial_x^2+\partial_y^2)a_2^t+2 mL(1+2 mL)a_2^t+B\left(\partial_x b-\partial_y a\right)=2L^3\left|c_0\right|^2 \left|\Psi_0^{\mathrm{lat}}\right|^2,
\end{eqnarray}
and the non-trivial Einstein equations coming from $G_{tr}$, $G_{tx}$, and $G_{ty}$ that are 
\begin{eqnarray}\label{backeomein}
\partial_x a+\partial_y b&=&0,\nonumber\\
\partial_x^2 b-\partial_{x}\partial_{y} a+2(2 m^2 L^2 + m L - 1)b-\frac{\sqrt{2}}{L}\partial_x a_2^t&=&-\frac{iL}{2}|c_0|^2\left(\Psi_0^{\mathrm{lat}}\partial_y{\Psi_0^{\mathrm{lat}}}^*-{\Psi_0^{\mathrm{lat}}}^*\partial_y{\Psi_0^{\mathrm{lat}}}\right),\\
\partial_y^2 a-\partial_{x}\partial_{y} b+2(2 m^2 L^2 + m L - 1)a+\frac{\sqrt{2}}{L}\partial_y a_2^t&=&-\frac{iL}{2}|c_0|^2\left(\Psi_0^{\mathrm{lat}}\partial_x{\Psi^{\mathrm{lat}}_{0}}^*-{\Psi^{\mathrm{lat}}_{0}}^*\partial_x{\Psi^{\mathrm{lat}}_{0}}-2iBy\left| \Psi_0^{\mathrm{lat}}\right|^2\right).\nonumber
\end{eqnarray}

For the rectangular lattice, we have a solution that is periodic in $x, y$ with periodicity $\frac{v_1}{\sqrt{B}}$ in the $x$ direction and $\frac{2\pi}{v_1\sqrt{B}}$ in the $y$ direction, therefore each of the functions can be expanded as a double Fourier series in $x,y$\footnote{Here we set $B=1$, which implies that each unit cell has a net flux density of $2\pi$.}
\begin{eqnarray}\label{ansatz}
f_i(x,y)=\sum\limits_{k,l}v_1 e^{2\pi i k\frac{x}{v_1}}e^{-il v_1 y}e^{-ikl\pi -\frac{k^2 \pi ^2}{v_1^2}-\frac{1}{4} l^2 v_1^2} \tilde{f_i}(k,l).
\end{eqnarray}
Using the Poisson summation formula, one can bring the Fourier transform of the source term into the form of the above equation. This trick helps us to reduce the coupled partial differential equations \eqref{backeommax} and \eqref{backeomein} to simple algebraic equations for the coefficients $\tilde{f_i}(k,l)$.

Plugging \eqref{ansatz} into the equations \eqref{backeommax} and \eqref{backeomein}, the algebraic equations for the $\tilde{f_i}(k,l)$ becomes
\begin{eqnarray}
\left(\frac{4 k^2\pi ^2}{v_1^2}+l^2 v_1^2-\sqrt{2} m-2 m^2\right)\tilde{a}_2^t -i \left(\frac{2k \pi }{v_1}\tilde{b}+l v_1 \tilde{a}\right)&=&-\frac{|c_0|^2}{4}\sqrt{\frac{2}{\pi }}\nonumber\\
2k\pi \tilde{a}-l v_1^2\tilde{b}&=&0\nonumber\\
2 k \pi l \tilde{b}+\left(l^2 v_1^2+2-\sqrt{2} m-2 m^2\right)\tilde{a}+2ilv_1\tilde{a}_2^t&=&\frac{i l v_1|c_0|^2}{4 \sqrt{2 \pi }}\nonumber\\
2 k \pi l \tilde{a}+\left(\frac{4k^2 \pi ^2}{v_1^2}+2-\sqrt{2} m-2 m^2\right)\tilde{b}+\frac{4 i k \pi }{v_1}\tilde{a}_2^t&=&\frac{ik|c_0|^2}{2v_1} \sqrt{\frac{\pi }{2}}.
\end{eqnarray}
The above equations show that $\tilde{a}$ and $\tilde{b}$ are pure imaginary and $\tilde{a}_2^t$ is a real function.
The solutions to these equations for $k=l=0$ are
\begin{eqnarray}
\tilde{a}=0,\hspace*{1cm}\tilde{b}=0,\hspace*{1cm}\tilde{a}_2^t=\frac{|c_0|^2}{2m\left(\sqrt{2}+2m\right)\sqrt{2 \pi }}
\end{eqnarray}
and in all other cases one finds
\begin{eqnarray}
\tilde{a}&=&\frac{il v_1^3 \left[4 k^2 \pi ^2+v_1^2 \left(4-\sqrt{2} m-2 m^2+l^2 v_1^2\right)\right]}{D}|c_0|^2\\ \tilde{b}&=&\frac{2\pi ikv_1 \left[4 k^2 \pi ^2+v_1^2 \left(4-\sqrt{2} m-2 m^2+l^2 v_1^2\right)\right]}{D}|c_0|^2,\nonumber\\
\tilde{a}_2^t&=&\frac{-12 k^2 \pi^2 v_1^2 + v_1^4 \left(-4 + 2 \sqrt{2} m + 4 m^2 - 3 l^2 v_1^2\right)}{D}|c_0|^2
\end{eqnarray}
where
\bea
D&=&4\sqrt{2 \pi }\Big[16 k^4 \pi ^4+8 k^2 \pi ^2 v_1^2 \left(l^2 v_1^2-\sqrt{2} m-2 m^2\right)\nonumber\\&+&v_1^4 \left(4 \sqrt{2} m^3+4 m^4+l^4 v_1^4-2 \sqrt{2} m \left(1+l^2 v_1^2\right)-2 m^2 \left(1+2 l^2 v_1^2\right)\right)\Big].
\eea
\subsection{Visualization of the Modulated phase}
\begin{figure}
\centering
\includegraphics[scale=.5]{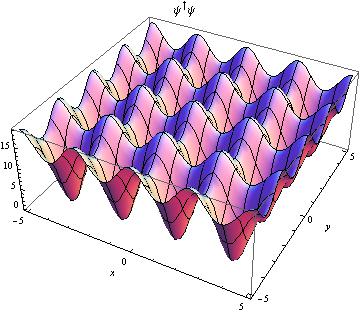}
\caption{The fermionic vortex lattice configuration for $m=\frac{1}{\sqrt{3}}$.}
\label{fig:psi}
\end{figure}

\begin{figure}
\centering
\begin{subfigure}
\centering
\includegraphics[scale=.5]{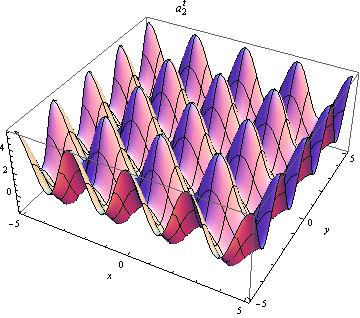}
\end{subfigure}
\begin{subfigure}
\centering
\includegraphics[scale=.5]{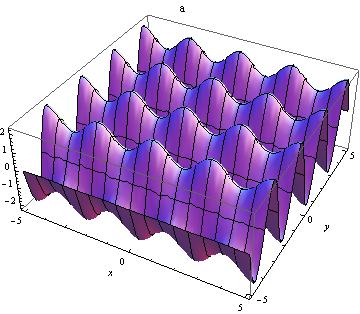}
\end{subfigure}
\caption{\textit{Left plot}: Flux configuration $a_2^t(x,y)$ for $m=\frac{1}{\sqrt{3}}$. \textit{Right plot}: Metric configuration $a(x,y)$ for $m=\frac{1}{\sqrt{3}}$.}
\label{fig:at-a}
\end{figure}

In this subsection we show different plots of the vortex lattice solution. In the Figure (\ref{fig:psi}) the fermionic lattice is plotted at order $\mathcal{O}(\epsilon)$ as a function of $(x,y)$. Figure (\ref{fig:at-a}) shows the spatially modulation of the temporal component of the gauge field $a_2^t(x,y)$ and metric $a(x,y)$.

\begin{figure}
\centering
\begin{subfigure}
\centering
\includegraphics[scale=.5]{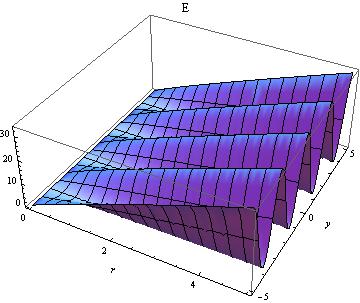}
\end{subfigure}
\begin{subfigure}
\centering
\includegraphics[scale=.5]{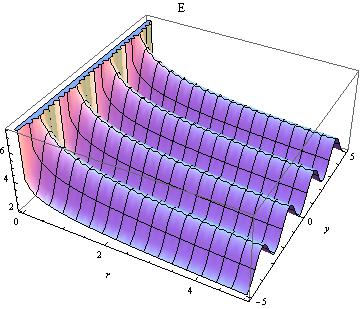}
\end{subfigure}
\caption{\textit{Left plot}: The electric field configuration for $m=\frac{1+\sqrt{3}}{2}$. \textit{Right plot}: The electric field configuration for $m=\frac{1}{\sqrt{3}}$.}
\label{fig:Ery}
\end{figure}

In the Figure (\ref{fig:Ery}) one can compare the profile of the electric field in the bulk at a constant $x$-slice as a function of $(r,y)$ for different mass parameters. The physical significance of these two plots is difference of the behavior of the electric field near the UV boundary. It is worth noting that in the case of alternative quantization, the electric field always reaches the UV boundary. 

In all plots, we consider $B=1$, $v_1=\sqrt{2\pi}$, and $c_0=4$.  Since the coefficients in the Fourier decomposition are exponentially suppressed as functions of $k^2$ and $l^2$, we have got a well approximation by running $k,l$ from $-5$ to $5$ (i.e. we have approximated the series with their first 121 terms).

\section{Discussion}
In this paper we have considered a magnetic AdS$_2\times\mathbb{R}^2$ background which is abruptly terminated in the IR. The background magnetic field breaks the translation symmetry along the $y$-direction. We considered Dirac fermions at order $\mathcal{O}(\epsilon)$ on this background. The fermions lie in Landau levels due to the magnetic field. Considering a specific superposition of the lowest Landau level solution, which are spin polarized, we have constructed a fermionic vortex lattice. Turning off the source term, we have solved the coupled PDE's of the metric and the gauge field sourced by the fermionic lattice. The back-reacted geometry comes to a crystalline structure. The spontaneously formed crystalline geometry supports an electric field and thus a finite charge density at $\mathcal{O}(\epsilon^2)$. The electric field can reach the boundary for specific range of parameters.

The lattice formation has several effects on the physics, including the thermodynamic functions. In order to compute corrections to the free energy and other thermodynamic quantities due to the lattice, explaining the role of $r_0$ is necessary. In the IR wall geometry, one can think of $r_0$ as a proxy for a confinement scale $\Lambda^{-1}$ in the confinement phase or a temperature $T^{-1}$ in the deconfined phase, which is represented by a horizon at some $r<r_0$ \cite{Bao:2013fda}. The leading corrections to the thermodynamic functions can be deduced from this correspondence.

The free energy in the field theory can be computed from the on-shell value of the bulk action and other observables can be derived from it. A simple dimensional analysis shows that in the standard quantization the free energy, entropy and specific heat densities take corrections as 
\begin{eqnarray*}\label{free}
\mathcal{F} &\sim& T+\epsilon^2 T^{-2m L}+\cdots\\
\mathcal{S} &\sim& 1+\epsilon^2 T^{-2m L-1}+\cdots\\
\mathcal{C} &\sim& \epsilon^2 T^{-2m L-1}+\cdots.
\end{eqnarray*}
We must note that the perturbation expansion is valid while $\epsilon \ll T^{ mL+1/2}$. In terms of the IR cut-off, $r_0$, this is equivalent to $\epsilon \ll r_0^{-mL-1/2}$. Considering $\epsilon$ as the distance away from the critical point, we see that decreasing $m$ extends the validity of linearised expansion region. For low temperatures, the above corrections become more important for the case of alternative quantization, where the sign of $m$ changed, while for high temperatures the converse is true.

\vspace{8mm}
It would be interesting to further explore this analytical fermionic vortex lattice in the following directions:

\begin{itemize}
\item
It is worth to generalize the fermionic vortex lattice to geometries with Lifshitz and/or hyperscaling violating exponents, specifically the case of $\eta$-geometries (the case where $z\to\infty$ and $-\theta/z=\eta$ is a constant). 

\item
The most important achievement of including the lattice effects in the dual field theory was the reconstruction of the Drude peak and a also reading the exponent of the power-law behavior in an intermediate frequency range of the optical conductivity. It would be interesting to compute the current-current correlators to study these features in this model. 

\item
A more natural setup to construct a vortex lattice is to consider a black-brane horizon in the IR, instead of a hard wall.

\item
We have discussed the lattice formation in this paper for standard and alternative quantizations of Dirac fermions. These are not the only possible quantizations. It would be interesting to investigate the effect of other possible quantizations, such as mixed quantization \cite{Laia:2011zn} in the lattice formation.

\item
We have only considered the lattice formation due to LLL solutions which are spin polarized. While the excited Landau levels ($n>0$) contain both spin-up and spin-down components, it is interesting to construct lattice solutions from the excited states. This can investigate the role of spin polarization in the lattice solution.
\end{itemize}

\section*{Acknowledgments}
We would like to thank  D. Tong, M.M. Sheikh-Jabbari, A. Vaezi, N. Bao, D. Allahbakhshi, and A. Naseh for useful discussions and comments. We would also like to thank M. Alishahiha, for fruitful discussions, comments and all of his supports during this work. We also acknowledge the use of M. Headrick's excellent Mathematica package "diffgeo".

\end{document}